\documentclass[
aps,
prc,
showpacs,
preprint,
superscriptaddress,
nofootinbib,
floatfix]
{revtex4}
\usepackage{graphicx,
longtable,
slashed}%

\newcommand{\beq}{\begin{equation}}
\newcommand{\eeq}{\end{equation}}
\newcommand{\bqa}{\begin{eqnarray}}
\newcommand{\eqa}{\end{eqnarray}}

\newcommand{\sump}{\mathop{{\sum\,}'}}

\begin{document}  

\title{A relation between isospin-symmetry breaking correction to superallowed beta decay 
and energy of charge-exchange giant monopole resonance}

\author{Vadim Rodin\footnote{Now at Stuttgart Technology Center, Sony Deutschland GmbH, D-70327, Stuttgart, Germany, e-mail: Vadim.Rodin@eu.sony.com}} 
\affiliation{Institut f\"{u}r Theoretische Physik der Universit\"{a}t  
T\"{u}bingen, Auf der Morgenstelle 14, D-72076 T\"{u}bingen, Germany} 
 
\date{\today}

\begin{abstract} 

After application of an analytical transformation, a new exact representation for the nuclear isospin-symmetry breaking correction $\delta_C$ to superallowed beta decay is obtained. 
The correction is shown to be 
essentially the reciprocal of the square of an energy parameter $\Omega_M$ which characterizes 
the charge-exchange monopole strength distribution.
The proportionality coefficient in this relation is determined by basic properties of the ground state of the even-even parent nucleus, and should be reliably calculable in any realistic nuclear model.
Therefore, the single parameter $\Omega_M$ contains all the information about the properties of excited $0^+$ states needed to describe $\delta_C$.
This parameter can possibly be determined experimentally by charge-exchange reactions. Basic quantities of interest are calculated within the  isospin-consistent continuum random phase approximation, and the values of $\delta_C$ are compared with the corresponding results from other approaches.

\end{abstract}

\pacs{
24.80.+y, 
23.40.Bw, 
23.40.Hc, 
21.60.-n, 
21.10.Sf, 
12.15.Hh  
}
\keywords{beta decay, isospin breaking} 
\maketitle 
 
\section{Introduction} 

Superallowed $0^{+} \rightarrow 0^{+}$ $\beta$ decays (SA $\beta$ decays) allow to test fundamental properties of the weak interaction, such as the Conserved Vector Current (CVC) hypothesis and the unitarity of the Cabibbo-Kobayashi-Maskawa (CKM) matrix $V_{ij}$ (see, e.g., a recent review~\cite{TH10} by Towner and Hardy).
The decay rates for SA Fermi transitions between $T=1$ nuclear multiplet states have accurately been measured in a dozen nuclei. Since the CVC hypothesis is only true in the isospin-symmetry limit,  an uncertainty enters in the analysis of the experimental $ft$-values depending on a model calculation of the effect of  isospin breaking in nuclei.
Although the breaking is weak,
the current situation is such that the theoretical uncertainties in calculated correction terms predominate over the experimental uncertainties in the SA $\beta$ decay data. This calls for a better accuracy of the theory applied to interpret the experimental results.

From the 2009 survey of experimental data, Hardy and Towner~\cite{HT09} determined
$|V_{ud}| = 0.97425 \pm 0.00022$, which, 
combined with the complimentary experimental data on $|V_{us}|$ and $|V_{ub}|$, gave:
\begin{equation}
|V_{ud}|^2+|V_{us}|^2+|V_{ub}|^2 = 0.99990 \pm 0.00060
\label{unit}
\end{equation}
for the norm of the first raw of the CKM matrix.
Thus, this test confirmed the unitarity of the CKM matrix with an accuracy of $0.06 \%$.

Isospin symmetry is slightly broken in nuclei, mainly by the Coulomb interaction. This leads to a small reduction of the nuclear matrix element $M_F$ for SA Fermi transitions between the ground state (g.s.) of the even-even parent nucleus and its isobaric analog state (IAS) in the odd-odd daughter nucleus:
\beq
\left| M_F\right|^2=\left| M^0_F\right|^2(1-\delta_C),
\label{deltaC}
\eeq
where $\left| M^0_F\right|^2\equiv 2T_0=|N-Z|$ is the exact-symmetry value,
with $T_0=|T_z|$ being the isospin 
of the g.s. of the even-even parent nucleus, and $\delta_C>0$ is the isospin-symmetry-breaking correction. 

There have been a number of methods used recently to calculate the correction $\delta_C$:
shell model with Saxon-Woods and Hartree-Fock radial functions \cite{HT09,TH08},
relativistic Hartree and Hartree-Fock approaches with the random phase approximation (RPA)~\cite{Li09}, an isovector monopole resonance (IVMR) model~\cite{auer09}, and self-consistent isospin- and angular-momentum-projected nuclear density functional theory~\cite{Sa11}. Still, there is a significant spread in the obtained values of $\delta_C$. Therefore, a better understanding of the aspects of nuclear structure that are important for more accurate evaluation of $\delta_C$ is needed.

The main purpose of this work is to derive a new exact representation for the correction $\delta_C$, which emphasizes the role of the physical charge-exchange monopole strength distributions, that can be probed experimentally. 
After application of an exact analytical transformation, $\delta_C$ is 
shown to be essentially reciprocal of the square of an energy parameter $\Omega_M$ which characterizes 
charge-exchange monopole strength distributions.
The proportionality coefficient in this relation is determined by basic properties of the ground state of the even-even parent nucleus, and should be reliably calculated in any realistic nuclear model.
Therefore, the single parameter $\Omega_M$ contains all the information about the properties of excited $0^+$ states needed to describe $\delta_C$.
The possibility of experimental determination of this parameter in charge-exchange reactions is discussed. Also in this paper
basic quantities of interest are calculated for few nuclei within the isospin-consistent continuum-RPA, and the obtained values of $\delta_C$ are compared with the corresponding results from other approaches.

\section{A new expression for the nuclear Coulomb correction $\delta_C$ }

In all the experimental cases of interest, the IAS of the g.s. 
$|0\rangle $ of a even-even parent nucleus with $T=1$ is an isolated 
low-lying state (or, in most of the cases, even the g.s.) in the daughter odd-odd nucleus. 
This physical IAS contains, along with the major $T=1$ component, also various small isospin admixtures. 

The representation for $\left| M_F\right|^2$ can identically be transformed as follows:
\bqa
\left| M_F\right|^2\equiv \langle 0|\hat T^{(+)} |IAS\rangle \langle IAS |\hat T^{(-)} |0\rangle 
&=& 2T_0 + S^{(+)}_F-S^{(-)}_F.
\label{1}
\eqa
Here, $S^{(+)}_F\equiv\sum\limits_{j} \left | \langle j | \hat T^{(+)} |0\rangle \right |^2$,
$S^{(-)}_F\equiv\sump\limits_{i} \left | \langle i | \hat T^{(-)} |0\rangle \right |^2$, 
and $\sump\limits_{i}\equiv\sum\limits_{i} - \sum\limits_{i=IAS}$ runs over all physical $0^+$ states, but the IAS, in the daughter nucleus, $\hat T^{-}=\sum_{a}\tau_a^{-}$ and $\hat T^{(+)}=\left(\hat T^{(-)}\right)^\dagger$ are the standard isospin lowering and raising operators, respectively. Note that Eq.~(\ref{1}) represents simply a version of the Ikeda sum rule for Fermi transitions: 
$\left| M_F\right|^2+S^{(-)}_F-S^{(+)}_F=2T_0=N-Z$. Hereafter we consider for definiteness the case $N>Z$, which can readily be generalized for the case $N<Z$.

Therefore, the Coulomb correction $\delta_C$ can be represented in the following form:
\beq 
\delta_C=\frac{1}{2T_0}(S^{(-)}_F-S^{(+)}_F)
\label{11}
\eeq

It proves very useful to further transform Eq.~(\ref{11}) to explicitly relate $\delta_C$ to isospin-breaking terms of the total nuclear Hamiltonian $\hat H$, which include the Coulomb interaction and the small isospin-violating part of the nuclear forces.
For this, one introduces auxiliary operators $\hat V^{(\mp)}_C\equiv \pm \left[\hat H,T^{(\mp)}\right]$ which are determined by these isospin-breaking terms, and one uses an exact relation between the matrix elements of $\hat V^{(\mp)}_C$ and $\hat T^{(\mp)}$:
\beq 
\langle s | \hat V_C^{(\mp)} |0\rangle=\omega_s \langle s | \hat T^{(\mp)} |0\rangle,
\label{12}
\eeq
with $\omega_s=E_s-E_0$ being the excitation energy of a state of the isobaric odd-odd daughter nucleus measured from the g.s. of the parent nucleus. The degree of fulfillment of Eq.~(\ref{12})  in a nuclear model can serve as an important check of the isospin consistency of the model. In particular, Eq.~(\ref{12}) ensures the equalities 
\beq
S^{(-)}_F={S}^{(-)}_{C[-2]}; \ \  S^{(+)}_F=S^{(+)}_{C[-2]},
\label{122}
\eeq
where ${S}^{(-)}_{C[L]}\equiv \sump\limits_{i} \left | \langle i | 
\hat V_C^{(-)} |0\rangle \right |^2\omega_i^L$ and $S^{(+)}_{C[L]}\equiv \sum\limits_{j} \left | \langle j | \hat V_C^{(+)} |0\rangle \right |^2\omega_j^L$ are the energy-weighted Coulomb sum rules.
As a result, 
one gets a representation equivalent to Eq.(\ref{11}):
\beq
\delta_C=\frac{1}{2T_0}({S}^{(-)}_{C[-2]}-S^{(+)}_{C[-2]})
\label{deltaC1}
\eeq
Although physically Eq.(\ref{11}) and Eq.(\ref{deltaC1}) are equivalent, it is preferable to use Eq.(\ref{deltaC1}) in a model calculation as this representation is much less sensitive to possible residual isospin inconsistencies of the model. 

The charge-dependent isospin-breaking interaction is dominated by the Coulomb interaction between protons.
Because the Coulomb force is of the long range, 
the one-body Coulomb mean field is mainly determining the transition operators $\hat V_C^{(\pm)}$ in Eq.(\ref{deltaC1}), which in this case also become one-body monopole charge-exchange operators.
Therefore, most of the strength for the transition operators is exhausted by the corresponding giant isovector charge-exchange monopole resonance (IVMR), associated with
the $2\hbar\omega$ particle-hole excitations of proton-neutron type with $J^\pi=0^+$.
The importance of the IVMR as a doorway state for the isospin mixing of the IAS was realized already in the early years of the IAS studies~\cite{auer72,auer83}, and was re-emphasized recently in Ref.~\cite{auer09}. 

Further, one can introduce an auxiliary energy $\Omega_M$ defined as 
\beq
\Omega_M^2\equiv\frac{{S}^{(-)}_{C[0]}-S^{(+)}_{C[0]}}{{S}^{(-)}_{C[-2]}-S^{(+)}_{C[-2]}}.
\label{Omega}
\eeq
This energy characterizes the charge-exchange monopole strength distributions in odd-odd isobaric nuclei. Then the original expression for $\delta_C$ (\ref{deltaC1}) can be identically rewritten as
$\delta_{c}=\frac{1}{2T_0} \frac{1}{\Omega_M^2} 
\left(\langle 0 | \left[\hat V_C^{(+)},\hat V_C^{(-)}\right ] |0\rangle -\omega_{A}^2 \left| M_F \right|^2 \right)$.
Here, we again have used $\sump\limits_{i}\equiv\sum\limits_{i} - \sum\limits_{i=IAS}$ and $\langle IAS | \hat V_C^{(-)} |0\rangle\equiv\omega_{A}  M_F $.
From this expression one obtains 
$
\left| M_F \right|^2=\frac{2T_0}{1-\frac{\omega_{A}^2}{\Omega_M^2}}(1-\frac{1}{2T_0\Omega_M^2} \langle 0 | \left[\hat V_C^{(+)},\hat V_C^{(-)}\right ] |0\rangle),
$
and, finally, arrives at the following expression:
\beq
\delta_C = 
\displaystyle\frac{1}{\Omega_M^2-\omega_{A}^2} 
\left(\frac{1}{2T_0}\langle 0 | \left[\hat V_C^{(+)},\hat V_C^{(-)}\right ] |0\rangle -\omega_{A}^2 \right)
\label{deltaC11}
\eeq

Thus, one sees that $\delta_C$ (\ref{deltaC11})
is determined by two energies, $\Omega_M$ and $\omega_{A}$, which are the only input in the problem related to the spectrum of $0^+$ states in the odd-odd daughter nuclei, and by the properties of the g.s. of the parent nucleus via the g.s. expectation value of the commutator $\langle 0 | \left[\hat V_C^{(+)},\hat V_C^{(-)}\right ] |0\rangle$.
Both  $\omega_{A}$ and $\Omega_M$ can be determined experimentally (the former is in fact already very accurately known; the value of the latter can be determined by charge-exchange reactions on the parent nucleus, see below). The numerical simulations (see below) indicate that a strong inequality $\Omega_M\gg\omega_{A}$ is fulfilled, that is to expect because of the high IVMR energy.

Now we would like to evaluate the expectation value 
$\langle 0 | \left[\hat V_C^{(+)},\hat V_C^{(-)}\right ] |0\rangle$ 
in the dominant mean field approximation, $\hat V_C=\sum_a U_C(r_a)(1-\tau_{az})/2$, with $U_C(r)$ being the Coulomb mean field. 
The realistic potential $U_C(r)$ resembles very much that of the uniformly charged sphere, which is a quadratic function: $U_C(r)=\displaystyle\frac{Ze^2}{2R}(3-(r/R)^2)$  inside a nucleus $ r\le R$, where $R$ is the nuclear radius. It turns out that if one extends this quadratic dependence also to the outer region  $r>R$ 
(instead of proportionality to $1/r$), this gives numerically just a small deviation in the Coulomb sum rules.
Thus, the Coulomb sum rules are determined by single-particle charge-exchange operators $\hat V_C^{(\mp)}\equiv \hat U_C^{(\mp)}=\sum_a U_C(r_a)\tau^\mp_a$, where the term in $U_C(r)$, proportional to $r^2$, gives the dominant contribution to the sum rules.
Further, one has
\beq
\left[\hat V_C^{(+)},\hat V_C^{(-)}\right]=\sum_a U_C^2(r_a) \tau_{az},
\label{dd}
\eeq
and finally gets
\beq
\frac{1}{2T_0}\langle 0 | \left[\hat V_C^{(+)},\hat V_C^{(-)}\right ] |0\rangle
=\frac{1}{2T_0}\int U_C^2(r)\varrho^{(-)}(r)
d^3r\equiv \overline{U_C^2},
\label{dd1}
\eeq
where the bar means averaging over the neutron excess density $\varrho^{(-)}(r)=\varrho_n(r)-\varrho_p(r)$ defined as the difference between the total neutron and proton number densities $\varrho_n(r)$ and $\varrho_p(r)$, respectively.

The nominator in Eq.~(\ref{deltaC11}) is subject to a strong cancellation between the two terms.
This can be best seen if one introduces the value of the Coulomb mean field averaged over the neutron excess density:
\beq
\overline{U_C} \equiv \frac{1}{2T_0}\int U_C(r)\varrho^{(-)}(r)d^3r. \label{ddd}
\eeq
By adding and subtracting $\overline{U_C}^2$ in the nominator of Eq.~(\ref{deltaC11}),
one gets:
\beq
\delta_C = \displaystyle\frac{1}{\Omega_M^2-\omega_{A}^2} 
\left(\overline{(U_C- \overline{U_C})^2}+(\overline{U_C}^2-\omega_{A}^2) \right).
\label{deltaC111}
\eeq
Each of the two terms in the nominator of Eq.~(\ref{deltaC111}) is now much smaller than its counterpart in Eq.~(\ref{deltaC11}):
$\overline{(U_C- \overline{U_C})^2}\ll\overline{U_C^2}$ as a consequence of the smoothness of the Coulomb mean field, 
and $(\overline{U_C}^2-\omega_{A}^2)\ll\omega_{A}^2$ since $\overline{U_C}$ provides a leading contribution to 
the Coulomb displacement energy 
(see, e.g.,~\cite{auer72,auer83}). 
The numerical simulations below show that only for light nuclei is the term $\overline{(U_C- \overline{U_C})^2}$ comparable with $\overline{U_C}^2-\omega_{A}^2$, for heavier systems the inequality  
$\overline{U_C}^2-\omega_{A}^2\gg \overline{(U_C- \overline{U_C})^2}$ holds. 

Different models must give similar results for $\overline{U_C^2}$ and $\overline{U_C}$ provided that the basic nuclear geometry (such as the mean radii of proton and neutron density distributions) can
reasonably be reproduced. 
Therefore, different values of $\delta_C$ obtained by different methods should mainly stem from
differences in corresponding values of $\Omega_M$.

Returning to the question of the possible experimental determination of $\Omega_M$, we note that 
the charge-exchange IVMR was first observed in pion single charge exchange reactions~\cite{pionexc,auer83a}. Recently, the IVMR has 
been studied in various charge-exchange reactions: ($^3$He,t)~\cite{Zegers00}, ($^3$He,tp)~\cite{Zegers00,Zegers03,Miki11} and (t,$^3$He)~\cite{Miki11}. The spin-flip IVMR was mainly excited in the experiments. Though the measurements are rather difficult to make, one may expect that the excitation of the non-spin-flip charge-exchange IVMR might be separated from its spin-flip counterpart (by means of polarized beams, or by comparing measurements at different projectile energies). 
Also, we note that the effective one-body transition operator leading to the IVMR excitation in charge-exchange forward-scattering reactions is determined by the $r^2$ dependence of the Bessel function 
$j_0(r)$~\cite{tad87}, in accord with the $r^2$ dependence of $U_C(r)$.
In such a case, $\Omega_M$ can directly be obtained from the experimental cross sections by a formula analogous to Eq.~(\ref{Omega}).

\section{Calculation results}

We consider here by way of example four decays: $^{10}$C$\to$$^{10}$B,$^{38}$K$\to$$^{38}$Ar, $^{66}$As$\to$ $^{66}$Ge, and $^{70}$Br$\to$$^{70}$Se. (The two latter cases, both with $A\approx 70$, are taken to check the abrupt drop in $\delta_C$ while going from $^{70}$Br to $^{66}$As as has appeared in calculations of Ref.~\cite{Li09}.)
We use here semi-phenomenological nuclear mean field and apply the continuum-RPA with Landau-Migdal zero-range forces to calculate the quantities of interest: $\omega_{A}$, $\overline{U_C^2}$ (\ref{dd1}), $\overline{U_C}$ (\ref{ddd}), $S^{(\pm)}_{F}$, $S^{(\pm)}_{C[0]}$, $S^{(\pm)}_{C[2]}$, $\Omega_M$~(\ref{Omega}), and finally $\delta_C$.

First calculations of the IVMR within the self-consistent HF + continuum-RPA approach were done in Ref.~\cite{AuerKlein83}. Here we use the relevant continuum-RPA equations from Refs.~\cite{Gor00,Gor01a}. Note, that we do not need any discretization of the single-particle (s.p.) continuum as done in Ref.~\cite{Li09}, because the equations are written in terms of the s.p. Green's functions.

The mean field is chosen as described in Ref.~\cite{Gor00}, and includes the fully phenomenological isoscalar part, with its parametrization tracing back to Chepurnov's potential~\cite{Chep67}, and both the symmetry potential and the mean Coulomb field calculated in the Hartree approximation. 
The chosen dimensionless intensity $f'=1.0$ of the isovector part of the Landau-Migdal forces 
determines the symmetry potential via the isospin selfconsistency condition~\cite{Gor01a}.
Thus, the mean Coulomb field is the only source of isospin breaking in the present model.

Since the nuclei in question are open-shell ones, one would in principle need to take into account the pairing correlations, and, better, to use the continuum-QRPA~\cite{Rod03,Rod08,Rod11} instead of the continuum-RPA. 
However, the continuum-QRPA calculations are much more time consuming, and, more importantly, one can easily argue that the effect of nucleon pairing on the quantities in question must be small (since the pairing gap is much smaller than $\Omega_M$). Also, a more modern choice of the isoscalar mean-field parameters of Ref.~\cite{Rod11}, which allows for their $A$-dependence, is not expected to markedly affect the results.

In Table~\ref{tab1} the calculated excitation energies $\omega_{A}$ and charge radii $r_c$ are listed along with the corresponding experimental data.
Table 1 also contains calculated values of $\overline{U_C^2}$ (\ref{dd1}) and $\overline{U_C}$ (\ref{ddd}), columns 6 and 7, respectively. 
The underestimate of the experimental IAS energy in the calculations for heavier nuclei reflects the Nolen-Schiffer anomaly~\cite{Nolen69,Auer69,Shlomo78}.

The fact that in lighter nuclei, in particular in the $A = 38$ case, the calculated Coulomb displacement energies are
larger than the experimental ones is apparently related to the global parametrization of the phenomenological mean field chosen in the paper, which was fixed to fit properties of medium-heavy and heavy nuclei, and can lead to larger deviations for light nuclei. It can also be seen in Table I that the calculated Coulomb radius is smaller then the experimental one. Trying to fit the latter by an appropriate choice of the nuclear radius of the mean field, one would get a smaller calculated Coulomb displacement energy.

\begin{table}[h]
\caption{In columns 2 and 3 the experimental and calculated excitation energies $\omega_{A}$, measured from the g.s. energy of the corresponding even-even nuclei in 
SA $\beta$-decays from column 1, are listed (the calculated RPA values are corrected for the neutron-proton mass difference). The calculated charge radii
$r_c$ of the even-even nuclei are given in column 5, and the only available experimental value for $^{38}$Ar is given in column 4. In columns 6 and 7 the values of $\overline{U_C^2}$ (\ref{dd1}) and $\overline{U_C}$ (\ref{ddd}), respectively, are listed. 
    \label{tab1}}
    \begin{tabular}{rcccccc}
\hline
& \multicolumn{2}{c} {$\omega_{A}$, MeV}
& \multicolumn{2}{c} {$r_c$, fm}
& $\overline{U_C^2}$ & $\overline{U_C}$\\
& exp & calc & exp & calc 
& MeV$^2$ & MeV \\[4pt]	

\hline
        $^{10}$C  $\to$ $^{10}$B  & -1.397 
& -1.66 
& -- & 2.69 
& ~-8.96 & -2.91\\
        $^{38}$K  $\to$ $^{38}$Ar & ~5.533 
& 5.57 
& 3.40 & 3.30 
 & ~55.83 & ~7.36\\
        $^{66}$As $\to$ $^{66}$Ge & ~9.609
& 8.97 
& -- & 
3.93   
 & 142.80 & 11.91\\
        $^{70}$Br $\to$ $^{70}$Se & 10.109
& 9.49 
& -- & 3.99
& 159.18 & 12.59\\
\hline
    \end{tabular}
\end{table}

The other calculated quantities of interest, which characterize the IVMR strength distributions, $S^{(\pm)}_{F}$, $S^{(\pm)}_{C[0]}$, $S^{(\pm)}_{C[2]}$, and $\Omega_M$~(\ref{Omega}), are listed in Table~\ref{tab2}. One can see a fairly good agreement between the corresponding entries in columns 2 and 4, and those in 3 and 5, respectively, in agreement with Eq.~(\ref{122}). This is a clear evidence of the isospin self-consistency of the applied continuum-RPA.
Finally, the isospin-symmetry-breaking correction $\delta_C$ is listed in  columns 9 (obtained directly from the RPA solution) and 10 (calculated from Eq.(\ref{deltaC111})).
Both ways of calculating $\delta_C$ agree well, again as a consequence of the isospin self-consistency of the applied continuum-RPA.

\begin{table}[h]
\centering
\caption{Calculated quantities characterizing the IVMR strength distributions. Calculated $S^{(\mp)}_F$ and $S^{(\mp)}_{C[-2]}$ are listed in  columns 2,3 and 4,5, respectively. In columns 6,7 values of $S^{\mp}_{C[0]}$ are given, and in column 8 the calculated energy $\Omega_M$~(\ref{Omega}) is listed. 
The isospin-symmetry-breaking correction $\delta_C$ is listed in  columns 9 (obtained directly from the RPA solution) and 10 (calculated from Eq.(\ref{deltaC111})).
    \label{tab2} }
\begin{tabular}{c c c c c c c c c c}
	\hline
Decay & 
\multicolumn{2}{c}{$S_F$ (\%)} & \multicolumn{2}{c}{$S_{C[-2]}$ (\%)} 
& \multicolumn{2}{c}{$S_{C[0]}$, MeV$^2$}
& $\Omega_M$, MeV & \multicolumn{2}{c}{$\delta_C$ (\%)}
\\[2pt]
& -- & + & -- & + & -- & + &  & RPA & (\ref{deltaC111}) \\[4pt]	
\hline
$^{10}$C  $\to$ $^{10}$B  
& 0.065 & 0.36 & 0.074 & 0.39  
& 1.01 & 1.47  & 12.13 
& 0.147
& 0.142\\

$^{38}$K  $\to$ $^{38}$Ar
& 2.16 & 1.33 & 2.18 & 1.33
& 26.95 & 9.35 & 45.38 
& 0.434 &  0.436\\

$^{66}$As $\to$ $^{66}$Ge 
& 7.38 & 5.44 & 7.43 & 5.43 
& 96.51 & 19.77 &  61.90
& 0.992 & 1.007 \\

$^{70}$Br $\to$ $^{70}$Se
& 7.83 & 5.92 & 7.89 & 5.90 
& 109.47 &  21.17 &  66.66
& 0.992 & 0.993 \\
\hline
\end{tabular}
\label{Tab2}
\end{table}

%

Apart from the decay $^{66}$As$\to$ $^{66}$Ge, the present results for $\delta_C$ are close to those of Ref.~\cite{Li09} calculated within the RH+RPA, and also are systematically smaller than those of Ref.~\cite{TH08}.
The corresponding value of $\delta_C$ for the decay $^{66}$As$\to$ $^{66}$Ge is pretty close to the one for $^{70}$Br$\to$$^{70}$Se, in contrast to the conclusion of Ref.~\cite{Li09}, but in qualitative accord with the small relative change in $\delta_C$ between these decays as observed in Ref.~\cite{TH08}.  

One can try to explain the difference between the shell model and the RPA results in terms of the difference in $\Omega_M$. The former approach uses the differences in radial single-particle wave functions of the neutron and proton with the same quantum numbers; i.e. it employs a pure mean-field picture. In this picture the collectivity of the IVMR is missing, and the effective $\Omega_M$ must be less than $\Omega_M$ of the continuum RPA. In the latter approach a collective IVMR is formed by the repulsive residual particle-hole interaction and is thereby shifted to higher excitation energy (see also similar arguments in Ref.~\cite{auer09}).

Physically, the collectivization of the IVMR results in both its energy shift to higher energy and a reduction of its Coulomb strength. However, these effects are not independent and are related via an energy-weighted sum rule. Namely the existence of such a relation allows one to relate $\delta_C$ exclusively to a single energy parameter $\Omega_M$  which characterizes the monopole strength distribution, see Eqs.~(\ref{Omega},\ref{deltaC11},\ref{deltaC111}).  Therefore, both effects of the IVMR collectivization can effectively be accumulated in a single energy parameter $\Omega_M$.

Note, that an estimate of the effect of the isospin splitting of the IVMR goes beyond the framework of the RPA.
The splitting effectively pushes the monopole strength to higher excitation energies, and it is expected that $\Omega_M$ will further slightly increase. This would lead to a corresponding decrease of $\delta_C$, bringing them closer to the estimates of Ref.~\cite{auer09}.

\section{Conclusions}

In the present work a new exact representation for the correction $\delta_C$ is derived in which the role of the physical charge-exchange monopole strength distributions is emphasized.
After application of an exact analytical transformation, $\delta_C$ is 
shown to be essentially the reciprocal of the square of an energy parameter $\Omega_M$ which characterizes 
charge-exchange monopole strength distributions.
The proportionality coefficient in this relation is determined by basic properties of the ground state of the even-even parent nucleus, and it should be reliably calculated in any realistic nuclear model.
The possibility of experimental determination of the parameter $\Omega_M$ in charge-exchange reactions is discussed. Also in this paper
basic quantities of interest are calculated for a few nuclei within the isospin-consistent continuum-RPA, and the obtained values of $\delta_C$ are compared with the corresponding results by other approaches.

\end{document}